\documentclass[a4paper,10pt]{article}
\begin{document}
\title{A Family of Exactly-Solvable Driven-Diffusive Systems in One-dimension}
\author{F H Jafarpour\footnote{Corresponding author's e-mail:farhad@ipm.ir} \, and P Khaki \\ \\
{\small Bu-Ali Sina University, Physics Department, Hamadan, Iran}}
\maketitle
\begin{abstract}
We introduce an exactly-solvable family of one-dimensional
driven-diffusive systems defined on a discrete lattice. We find the
quadratic algebra of this family which has an infinite-dimensional
representation. We discuss the phase diagram of the system in a
couple of special cases.
\end{abstract}
\maketitle

One-dimensional driven-diffusive systems are systems of classical
particles with hard-core interactions moving in a preferred
direction which can be used to model many systems such as ribosomes
moving along a m-RNA, ions diffusing in a narrow channel under the
influence of an electric field or even cars proceeding on a long
road \cite{SCH,MGP,CSS}. These systems are usually defined on an
open lattice coupled with two reservoirs at both ends or on a
lattice with periodic boundary conditions. In long time limit the
system settles into an non-equilibrium steady state characterized by
some bulk density and the corresponding particle current. The
out-of-equilibrium steady state properties of these systems are not
only affected by the boundaries but also to the localized
inhomogeneities which play a crucial role not comparable to that in
classical equilibrium systems. Driven-diffusive systems have been
studied extensively in the past decade because of their unique
non-equilibrium properties.

There are different approaches to study the steady state of these
systems. The Matrix Product Formalism (MPF) is one of the most
powerful techniques in this field \cite{BE,DEHP,KS}. According to
this formalism one assigns an operator to every state of a lattice
site of the system. Now every configuration of the system is
associated with a product of such operators. The steady state
probability of such configuration is then given by a trace (for the
systems defined on a ring geometry) or a matrix element (for the
systems defined on an open lattice) of such products. For instance
for a three-states system on a lattice with periodic boundary
condition we define three operators, let us say $A$, $B$ and $E$,
associated with three different states of each lattice site. The
steady state weight of a given configuration similar to $ABEEABEAA$
is then proportional to $Tr[ABEEABEAA]$. Requiring that the
probability distribution function of the system defined above is
stationary provides us with an algebra of operators. For the systems
with nearest neighbors interactions it turns out that the algebra is
quadratic. In order to calculate the steady state weights one can
either work with the commutation relation of the operators or find a
representation for the associated quadratic algebra of the system
(for a recent review on the MPF see \cite{BE}).

So far only a couple of one-dimensional driven-diffusive models have
been solved exactly using the MPF \cite{BE}. Therefore it would be
interesting to investigate whether or not there are other models
which can be solved exactly using this technique. In this paper we
introduce a large family of one-dimensional driven-diffusive models
which under some constraints on the reaction rates can be solved
exactly i.e. the steady state weight of any configuration can be
calculated rigourously using the MPF. From there the mean values of
physical quantities such as the currents of particles and also their
concentrations can be obtained exactly. This three-states family
belongs to the systems with non-conserving dynamics and
nearest-neighbors interactions. These states belong to two different
types of particles and holes on each lattice site. It is not clear
that the steady state weights of such a system can generally be
written as matrix product states; however, we will present a
constraint under which it will possible to write these weights using
the MPF. In this regard, we will relate the quadratic algebra of our
system to the quadratic algebra of the systems with known
representations. A couple of members of this family of systems have
already been studied in literature. At the end of this paper we will
briefly discuss some other special cases which have not been studied
before.

During last decade several quadratic algebras have been introduced
and used to study the critical behaviors of one-dimensional
driven-diffusive systems. It has also been tried to classify certain
quadratic algebras and find their representations
\cite{IPR,ADR,AHR1}. One of the most well-known quadratic algebras
belongs to the Partially Asymmetric Simple Exclusion Process (PASEP)
in which the classical particles hop to the left and to the right
with the rates $p$ and $q$ on an open one-dimensional lattice of
length $L$. The jumps of particles are only successful provided that
the target sites are empty. The particles are injected into the
system from the left boundary with the rate $\alpha$ provided that
the first site of the lattice is empty. The particles are also
extracted from the last site of the lattice provided that it is
already occupied. The particles are subjected to the hardcore
interactions so that two particles cannot occupy a single site of
the lattice simultaneously. It is known that the steady state
weights of the PASEP can be written in terms of products of
infinite-dimensional square matrices which satisfy a quadratic
algebra of the following form \cite{DEHP,AHR2,SM}
\begin{equation}
\label{ALG1}
\begin{array}{c}
p AB-q BA=A+B\\
\alpha A \vert V \rangle =\vert V \rangle\\
\beta \langle W \vert B = \langle W \vert
\end{array}
\end{equation}
in which the operators $A$ and $B$ are associated with two different
states of each lattice site i.e. a particle and an empty site
respectively. It is known that this algebra has an
infinite-dimensional matrix representation for $p \ge q$ given by
the following matrices \cite{DEHP}
\begin{equation}
\label{REP1}
\begin{array}{c}
A=\frac{1}{p-q}\left(
    \begin{array}{ccccc}
      1-a & d_1 & 0 & 0 & \cdots \\
      0 & 1-a(\frac{q}{p}) & d_2 & 0 &  \\
      0 & 0 & 1-a(\frac{q}{p})^2 & d_3 &  \\
      0 & 0 & 0 & 1-a(\frac{q}{p})^3 &  \\
      \vdots &   &   &   & \ddots \\
    \end{array}
  \right), \\
B=\frac{1}{p-q}\left(
    \begin{array}{ccccc}
      1-b & 0 & 0 & 0 & \cdots \\
      d_1 & 1-b(\frac{q}{p}) & 0 & 0 &  \\
      0 & d_2 & 1-b(\frac{q}{p})^2 & 0 &  \\
      0 & 0 & d_3 & 1-b(\frac{q}{p})^3 &  \\
      \vdots &   &   &   & \ddots \\
    \end{array}
  \right)
  \end{array}
\end{equation}
and vectors
\begin{equation}
\label{REP2} \vert V \rangle=\left(
                                    \begin{array}{c}
                                      1 \\
                                      0 \\
                                      0 \\
                                      0 \\
                                      \vdots \\
                                    \end{array}
                                  \right),
\langle W \vert =\left(
                                    \begin{array}{ccccc}
                                      1&0&0&0&\cdots \\
                                    \end{array}
                                  \right)
\end{equation}
in which we have defined $a=1-\frac{p-q}{\alpha}$,
$b=1-\frac{p-q}{\beta}$ and
${d_i}^2=(1-(\frac{q}{p})^i)(1-ab(\frac{q}{p})^{i-1})$. The matrix
representation of (\ref{ALG1}) given by (\ref{REP1}) and
(\ref{REP2}) is quit well-defined in a sense that product of any
number of these matrices sandwiched between $\langle W \vert$ and
$\vert V \rangle$ is finite. In what follows we will look for those
systems which their steady states in terms of the MPF are given by
(\ref{ALG1}) and its representation i.e. (\ref{REP1}) and
(\ref{REP2}).

Let us define a new operator $E=\omega \vert V \rangle \langle W
\vert$ in which $\omega$ is a real number. By multiplying $\langle W
\vert\omega$ from the right in the second row and also $\omega\vert
V \rangle$ from the left in the third row of (\ref{ALG1}) one finds
\begin{equation}
\label{ALG2}
\begin{array}{c}
p AB-q BA=A+B\\
\alpha A E = E\\
\beta E B = E
\end{array}
\end{equation}
which was first introduced in \cite{DJLS}. Obviously this algebra
has the same representation given by (\ref{REP1}) and (\ref{REP2})
and one should only define $E$ as
\begin{equation}
\label{E} E=\omega \vert V \rangle \langle W \vert =\left(
    \begin{array}{ccccc}
      \omega & 0& 0 & 0 & \cdots \\
      0 & 0& 0 & 0 &  \\
      0 & 0& 0 & 0 &  \\
      0 & 0& 0 & 0 & \\
      \vdots &   &   &   & \ddots \\
    \end{array}
  \right).
\end{equation}
It can easily be verified that the operator $E$ has the property
$E^2=\omega E$ and also can be added to (\ref{ALG2}) without
changing its matrix representation to make a larger and well-defined
quadratic algebra. Therefore the following quadratic algebra has the
same matrix representation given by (\ref{REP1}), (\ref{REP2}) and
(\ref{E})
\begin{equation}
\label{ALG3}
\begin{array}{c}
p AB-q BA=A+B\\
\alpha A E = E\\
\beta E B = E\\
E^2=\omega E.
\end{array}
\end{equation}
The question is now whether or not one can find a three-states
model, defined on a lattice with a ring geometry, with a steady
state described by (\ref{ALG3}). By applying the standard MPF
\cite{BE,KS} we have found that the steady state probability
distribution function of the following three-states system
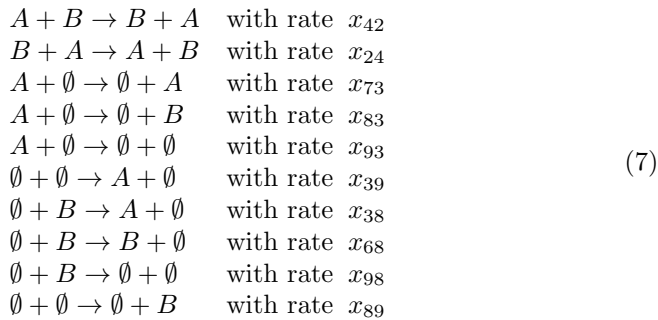
\begin{equation}
\label{Rules}
\begin{array}{lll}
A+B \rightarrow B+A & \mbox{with rate} \; \; x_{42} \\
B+A \rightarrow A+B & \mbox{with rate} \; \; x_{24} \\
A+\emptyset \rightarrow \emptyset+A & \mbox{with rate} \; \; x_{73} \\
A+\emptyset \rightarrow \emptyset+B & \mbox{with rate} \; \; x_{83} \\
A+\emptyset \rightarrow \emptyset+\emptyset & \mbox{with rate} \; \; x_{93} \\
\emptyset+\emptyset \rightarrow A+\emptyset & \mbox{with rate} \; \; x_{39} \\
\emptyset+B \rightarrow A+\emptyset & \mbox{with rate} \; \; x_{38} \\
\emptyset+B \rightarrow B+\emptyset & \mbox{with rate} \; \; x_{68} \\
\emptyset+B \rightarrow \emptyset+\emptyset & \mbox{with rate} \; \; x_{98} \\
\emptyset+\emptyset \rightarrow \emptyset+B & \mbox{with rate} \; \; x_{89} \\
\end{array}
\end{equation}
in which $A$, $B$ and $E$ can be associated with two particles of
different types and an empty site respectively, can be described by
(\ref{ALG3}) provided that we define
\begin{equation}
\label{Para}
\begin{array}{l}
p=x_{42}\\
q=x_{24}\\
\alpha=x_{73}\\
\beta=x_{68}\\
\omega=\frac{1}{x_{39}+x_{89}}(\frac{x_{93}}{x_{73}}+\frac{x_{98}}{x_{68}})
\end{array}
\end{equation}
and require the parameters to satisfy the following constraint
\begin{equation}
\label{Cons}
x_{39}(\frac{x_{83}}{x_{73}}-\frac{x_{38}+x_{98}}{x_{68}})=
x_{89}(\frac{x_{38}}{x_{68}}-\frac{x_{93}+x_{83}}{x_{73}}).
\end{equation}
This constraint together with the definitions (\ref{Para}) guarantee
that our model defined by (\ref{Rules}) has a well-defined algebra
given by (\ref{ALG3}) and that its steady state weights can be
expressed in terms of a matrix product states.

As we mentioned, a couple of special cases of this family of models
have already been studied in literature. For instance in \cite{EKLM}
the authors have studied an exactly solvable model with the
constraints $x_{83}=x_{38}=0$,
$x_{42}=\frac{1}{q}x_{24}=x_{73}=x_{68}=x_{39}=x_{89}=1$ and
$x_{93}=x_{98}=\omega$. In this case the particle number for both
$A$ and $B$ particles changes because of creation and annihilation
of them. They have found that the system undergoes a continuous
phase transition from a fluid phase to a maximal current phase for
$q<1$ by varying $\omega$ . For $q>1$ the system is always phase
separated and there is no phase transition. In another paper the
authors have considered an special case in which
$x_{24}=x_{83}=x_{38}=x_{98}=x_{89}=0$,
$x_{42}=x_{68}=x_{73}=x_{39}=1$ and $x_{93}=\omega$ \cite{ JG1}. In
this case the number of $B$ particles is conserved while the number
of $A$ particles can change due to the creation and annihilation. It
has been shown that, in this case where a finite density of $B$
particles exists on the ring, the model can be solved exactly and
undergoes a second-order phase transition by varying $\omega$ from a
phase in which the density of the empty sites is zero to another
phase where the density of the empty sites is nonzero. In \cite{JG2}
the authors have studied the same process with more general reaction
rates as $x_{24}=x_{83}=x_{38}=x_{98}=x_{89}=0$,
$x_{42}=\frac{1}{\beta}x_{68}=\frac{1}{\alpha}x_{73}=x_{39}=1$ and
$x_{93}=\omega$; however, in this case they assume that there is
only a single $B$ particle in the system. They have found that the
phase transition is now discontinuous and that shocks might appear
in the system at the transition point.

In what follows we consider the most general case where most of the
parameters in (\ref{Rules}) are nonzero which has not been studied
in literature yet. We first study the phase diagram of the model for
$x_{24}=0$ and $x_{42}=1$ \footnote{By resealing the time one can
always take one of the parameters equal to one.}. The partition
function of the system defined as sum of the weights of all
accessible configurations with at least one empty site, is given by
\begin{equation}
\label{PF} Z_{L}(\alpha,\beta,\omega,\xi)=Tr[(\xi A+B+E)^L]-Tr[(\xi
A+B)^L].
\end{equation}
The fugacity of $A$ particles $\xi$ is an auxiliary parameter and
the reason we have defined it will be clear shortly. It turns out
that the generating function of this partition function can be
calculated exactly. After some straightforward calculations one
finds
\begin{equation}
\label{GF}
G(\alpha,\beta,\omega,\xi,\lambda)=\sum_{L=1}^{\infty}\lambda^{L}
Z_{L}=\frac{\omega \lambda \frac{\partial}{\partial
\lambda}U}{1-\omega U}
\end{equation}
in which we have defined
$$U(\alpha,\beta,\xi,\lambda)=\sum_{L=0}^{\infty}\lambda^{L+1}\langle
W \vert (\xi A+B)^L \vert V \rangle.$$ It turns out that
\begin{equation}
U(\alpha,\beta,\xi,\lambda)=\frac{4\lambda}{f^{-}(\alpha)f^{+}(\beta)}
\end{equation}
where
$$f^{\pm}(x)=\frac{1}{x}(1-2x\pm\lambda(1-\xi)-\sqrt{(1+\lambda(1-\xi))^2-4\lambda}).$$
The phase diagram of the system can now be obtained by studying the
singularities of the generating function (\ref{GF}) for $\xi=1$. One
can easily see that in this case the generating function has two
different kinds of singularities: a square root singularity
$\lambda^{*}=\frac{1}{4}$ and two simple pole singularities which
come from denominator of (\ref{GF}) by solving the equation
$1-\omega U(\alpha,\beta,\xi=1,\lambda^{*})=0$. However, analyzing
the absolute values of singularities shows that the system can only
have two phases: a maximal current phase which is specified by the
square root singularity (for $\alpha+\beta > 1$ and
$(2-\frac{1}{\alpha})(2-\frac{1}{\beta})\geq \omega$) and a fluid
phase which is specified by the simple pole singularity (for
$\alpha+\beta < 1$ or $\alpha+\beta
> 1$ and $(2-\frac{1}{\alpha})(2-\frac{1}{\beta}) < \omega$). Keeping
$\xi=1$ the total density of the empty sites $\rho_{E}$ can easily
be calculated using
\begin{equation}
\rho_{E}=\lim_{L\rightarrow
\infty}\frac{\omega}{L}\frac{\partial}{\partial \omega} \ln Z_{L}.
\end{equation}
In the thermodynamic limit we have
$\rho_{E}\sim-\omega\frac{\partial \ln \lambda^{*}}{\partial
\omega}$. It turns out that the density of the empty sites is zero
in the maximal current phase while it is nonzero in the fluid phase.
The total density of $A$ and $B$ particles in each phase can also be
calculated exactly. Since the density of the empty sites is known
only one of these densities is independent. The auxiliary fugacity
$\xi$ can now help us find the density of $A$ particles as
\begin{equation}
\rho_{A}=\lim_{L\rightarrow
\infty}\frac{\xi}{L}\frac{\partial}{\partial \xi} \ln Z_{L}
\mid_{\xi=1}.
\end{equation}
The density of the $B$ particles is then
$\rho_{B}=1-\rho_{E}-\rho_{A}$. We have found that in the maximal
current phase $\rho_{A}=\rho_{B}=\frac{1}{2}$. In the fluid phase
both $\rho_{A}$ and $\rho_{B}$ are complicated functions of
$\alpha$, $\beta$ and $\omega$ and will not be presented here. The
particle currents for both species can also be calculated exactly in
this case. We have found that the current of $A$ particles $J_{A}$
is always equal to that of $B$ particles $J_{B}$. Our calculations
also show that the particle current is given by
$J_{A}=J_{B}=\frac{Z_{L-1}}{Z_{L}}$ which is equal to $\lambda^{*}$
in the thermodynamic limit. In the maximal current phase we simply
find them to be equal to $\frac{1}{4}$. This is actually in contrast
with the case studied in \cite{JG1} where the number of $B$
particles in one species is conserved. It has been shown that in
this case the currents can be different. For $x_{24}=q$, $x_{42}=1$,
$\alpha=1$ and $\beta=1$ the results are exactly those obtained in
\cite{EKLM}. On a lattice with periodic boundary conditions we have
found that the phase diagram of the model does not change even for
arbitrary $\alpha$ and $\beta$; therefore, we will not discuss this
case here.

One should note that (\ref{ALG3}) can also explain the steady state
of a system with open boundaries and two species of particles. The
particles of type $A$ ($B$) are injected from the left (right)
boundary with rate $\frac{1}{\omega}$ ($\frac{1}{\omega}$) and
extracted from the right (left) boundary with rate $\alpha$
($\beta$). All of the processes in (\ref{Rules}) might also take
place on the lattice. The partition function of the model for
$x_{24}=0$ and $x_{42}=1$ can also be calculated exactly and is
given by
\begin{equation}
Z_{L}(\alpha,\beta,\omega)=\langle W \vert (A+B+E)^L \vert V
\rangle=\sum_{i=1}^{L}\frac{i(2L-i-1)!}{L!(L-i)!}
\frac{\tilde{\alpha}^{-i-1}-\tilde{\beta}^{-i-1}}{\tilde{\alpha}^{-1}-\tilde{\beta}^{-1}}
\end{equation}
in which
\begin{equation}
\tilde{\alpha}=\frac{\alpha}{1+\alpha\omega\lambda_1},\tilde{\beta}=\frac{\beta}{1+\beta\omega\lambda_2}
\end{equation}
and
\begin{eqnarray}
\lambda_1=\frac{1}{2\alpha\beta\omega}(\alpha-\beta+\alpha\beta\omega-
\sqrt{(\alpha-\beta+\alpha\beta\omega)^2+4\alpha\beta^2\omega(1-\alpha)}),\\
\lambda_2=\frac{-1}{2\alpha\beta\omega}(\alpha-\beta-\alpha\beta\omega-
\sqrt{(\alpha-\beta+\alpha\beta\omega)^2+4\alpha\beta^2\omega(1-\alpha)}).
\end{eqnarray}
The phase diagram structure of the model can now be obtained by
studying the thermodynamic behavior of the partition function.
Equivalently one can study the zeros of this partition function as a
function of $\alpha$, $\beta$ or $\omega$. It turns out that the
system has again two different phases. The phase transition occurs
for $\alpha+\beta>1$ at
$\omega=(2-\frac{1}{\alpha})(2-\frac{1}{\beta})$ similar to the ring
geometry case. Both phases are symmetric that is the currents of
particles of different types are always equal.

In this paper we have investigated a general quadratic algebra
(\ref{ALG3}) associated with a family of exactly solvable
three-states reaction-diffusion systems with non-conserving dynamics
defined on a one-dimensional lattice with ring geometry. This
algebra is in fact the quadratic algebra associated with the PASEP
given by (\ref{ALG1}) besides the relation $E^2=\omega E$ which does
not change the representation of the algebra but it generates a new
algebra which allows us to study a new family of three-states
processes using the MPF. This family of three-states processes are
defined by ten nonzero reaction rates given by (\ref{Rules}) which
should satisfy a constraint given by (\ref{Cons}). Under this
constraint the steady state of the system can be written as a matrix
product form. We have considered the most general model of this type
on a lattice with periodic boundary conditions and studied its phase
diagram. As we have also mentioned, the generalized algebra
(\ref{ALG3}) can explain the steady state of a three-states system
with the reactions defined by (\ref{Rules}) but this time under open
boundary conditions where the particles are allowed to enter and
leave the lattice with some specific injection and extraction rates.
The phase diagram of the model has also been studied and the
partition function calculated exactly. Our approach can be
generalized and applied to other models similar to the model studied
in \cite{J} (which is a $p$-species model defined on a lattice with
periodic boundary conditions) to find a $p+1$-species exactly
solvable model. The results will be published elsewhere.

\end{document}